\documentclass[10pt,conference]{IEEEtran}
\IEEEoverridecommandlockouts

\usepackage{cite}
\usepackage{amsmath,amssymb,amsfonts}
\usepackage{algorithmic}
\usepackage{graphicx}
\usepackage{textcomp}
\usepackage{xcolor}
\usepackage{placeins}

\usepackage{tabularx}
\usepackage{array}

\def\BibTeX{{\rm B\kern-.05em{\sc i\kern-.025em b}\kern-.08em
    T\kern-.1667em\lower.7ex\hbox{E}\kern-.125emX}}
\begin{document}

\title{Adversarial attacks on hybrid classical-quantum Deep Learning models for Histopathological Cancer Detection\\


}

\author{\IEEEauthorblockN{Biswaraj Baral}
\IEEEauthorblockA{\textit{Quantum Computing Group} \\
\textit{Qausal AI}\\
San Ramon, CA, USA \\
biswa@qausal.ai}
\and
\IEEEauthorblockN{Reek Majumdar}
\IEEEauthorblockA{\textit{ Civil Engineering Department} \\
\textit{Clemson University}\\
Clemson, SC, USA \\
rmajumd@clemson.edu}
\and
 
\IEEEauthorblockN{Bhavika Bhalgamiya}
\IEEEauthorblockA{\textit{Dept of Physics and Astronomy}\\
\textit{Mississippi State University } \\
MS, USA\\
bgb182@msstate.edu}
\and
\IEEEauthorblockN{Taposh Dutta Roy}
\IEEEauthorblockA{\textit{Quantum Computing Group} \\
\textit{SVQC}\\
San Ramon, CA, USA \\
taposh@qausal.ai}
}
\maketitle

\begin{abstract}

We present an effective application of quantum machine learning in histopathological cancer detection. The study here emphasizes two primary applications of hybrid classical-quantum Deep Learning models. The first application is to build a classification model for histopathological cancer detection using the quantum transfer learning strategy. The second application is to test the performance of this model for various adversarial attacks. Rather than using a single transfer learning model, the hybrid classical-quantum models are tested using multiple transfer learning models, especially ResNet18, VGG-16, Inception-v3, and AlexNet as feature extractors and integrate it with several quantum circuit-based variational quantum circuits (VQC) with high expressibility. As a result, we provide a comparative analysis of classical models and hybrid classical-quantum transfer learning models for histopathological cancer detection under several adversarial attacks. We compared the performance accuracy of the classical model with the hybrid classical-quantum model using pennylane default quantum simulator. We also observed that for histopathological cancer detection under several adversarial attacks, Hybrid Classical-Quantum (HCQ) models provided better accuracy than classical image classification models. 
\end{abstract}


\section{Introduction}

Predictive models face the reality of encountering various attacks from vindictive entities. Adversarial attacks are one of these attacks, which mainly target AI models like Deep Learning (DL) or Machine Learning (ML) models. These attacks involve deliberately perturbing original input images with a carefully crafted noisy image, resulting in incorrect image classification by the model. Perturbed Images are imperceptible to the human eye, but it confuses the model, leading to misclassification. As per a recent study \cite{dong2023adversarial}, adversarial machine learning is a critical aspect of the ML field, pressing the need for practitioners and researchers to acknowledge and address the potential threats posed by adversarial attacks to the effectiveness and trustworthiness \cite{barreno2006can} of machine learning models. The use of machine learning in the healthcare system is increasing to make the diagnosis and decision system robust. Due to the widespread use of machine learning models in healthcare systems, such systems are at a high risk of adversarial attacks. One common impact of adversarial attacks that could be experienced in the healthcare system is misleading the insurance approval system. Insurance companies use predictive models to confirm the approval of insurance reimbursement. Fraudsters may integrate the insurance data with perturbed data and lead to false insurance claims \cite{AdversarialinHealth}. It is crucial for next-generation DL models to mitigate these attacks to solve the image misclassification problem.

In this study, we aim to  investigate the impact of adversarial attacks \cite{hirano2021universal} on classical Deep Learning (C-DL) ,and hybrid classical-quantum Deep Learning(HCQ-DL) models. The primary goal here is to present more resilient HCQ-DL models compared to C-DL models, which can  perform better during adversarial attacks in order to obtain better performance accuracy. Our HCQ-DL models are trained with quantum simulators. As a result, we provide a comparative study of C-DL and HCQ-DL models for histopathological adversarial images.
\\

The paper is organized as follows.  The details of model creation, QNN layer integration and generation of adversarial images using different types of adversarial attack algorithms used in are discussed in Section III. Results obtained from our experiment on different classical and hybrid classical quantum models are included in Section IV. The experiment performed in this study is summarized in conclusion Section V.

\section{Literature Review}
The recent development in computing technology and the availability of superior computing power and GPU leads to the extensive use of machine learning models in different sectors. Due to the availability of a large collection of health datasets, machine learning models are widely used in health sectors like gastroenterology, ophthalmology, pathology and dermatology for appropriate diagnosis and decision-making. Research performed on adversaries by different researchers in various applications of machine learning proved that almost all deployed machine learning models are extremely vulnerable to adversarial attacks. Formally the term `adversarial input' was first described in 2004 by Dalvi et al. when they designed the framework to defend different adversarial manipulation by spammers on spam classifier \cite{firstAdversarialTErm}. 

Finlayson et al. \cite{finlayson2018adversarial} used different white-box and block-box attacks to generate adversarial perturbations. The experiment was performed on three use cases of medical images classification: fundoscopy, chest x-ray and dermoscopy. The attack success rate upto $100\% $ with a confidence score of $100\%$ is achieved from the experiment. The adversarial experiment performed on a real-time smart healthcare system deteriorated the performance of the system \cite{newaz2020adversarial}. The experiment used 4 different black-box and white-box attack methods to generate adversarial perturbations. There was a significant drop in classification accuracy under both targeted and untargeted attacks. The highest success rate achieved under adversarial attacks is $15.68\%$. The adversarial experiment performed on ISIC dataset shows that there is a huge difference between the classification accuracy with and without adversarial perturbations \cite{isicDataset}. Selvakkumar et al. used a pre-trained vgg19 transfer learning model for binary image classification. The study used Fast Gradient Sign Method(FGSM) algorithm for adversarial image generation which drop the accuracy of classification from $88\%$ to $11\%$. 

There are several techniques that can be used for \textbf{adversarial attacks} on machine learning models. These threat models are categorized into black-box and white-box adversarial attack methods. In white-box attack the attacker has the information of deployed model like inputs, the architecture of model, internal gradients and weights and other parameters while in black-box the attacker has no access to such parameters. Some of the most common techniques of adversarial attacks are listed in the TABLE \ref{tab:attacks_examples} below.

\begin{table}[htpb]
\centering
\caption{Examples of commonly used adversarial attack techniques}
\begin{tabular}{|l|l|}
\hline
\textbf{Adversarial Technique} & \textbf{Examples of attacks}  \\ \hline

Gradient-based Attacks   & Auto Projected Gradient - \\& Descent (Auto-PGD) \cite{croce2020reliable}\\ & 
Fast Gradient Method \cite{goodfellow2014explaining}\\ &
Shadow Attack \cite{ghiasi2020breaking} "\\ \hline

Genetic Algorithms & Wasserstein Attack \cite{wong2019wasserstein}\\ &
Targeted Universal Adversarial - \\ & Perturbations \cite{hirano2020simple} \\ \hline

Boundary Attack & Brendel - Bethge Attack \cite{brendel2019accurate}\\ &
Threshold Attack\cite{thresholdAttack} \\& DeeFool Attack\cite{DeepFoolOriginal}  \\ \hline

Zeroth-order Optimization & Zeroth Order Optimisation (ZOO) \cite{chen2017zoo} \\ \hline

Transferability Attacks & Functionally Equivalent Extraction \cite{jagielski2020high} \\ &
Copycat CNN \cite{correia2018copycat} \\&
KnockoffNets \cite{orekondy2019knockoff}  \\ \hline

\end{tabular}
\label{tab:attacks_examples}
\end{table}

In this experiment, we attempt to generate adversarial perturbations using some of the methods of gradient-based attacks and boundary attacks only.\\ 
\textbf{Gradient-based attacks}: 
In gradient-based attacks, an attacker computes the gradients of the model with respect to the input data and then modifies the input data to maximize the loss function. This can be achieved using techniques such as the fast gradient sign method, the projected gradient descent method, or the momentum iterative method.\\
\textbf{Boundary attack}: In boundary attack-based attacks, an attacker generates a series of inputs that lie near the decision boundary of the model, and then perturbs these inputs in such a way that they are misclassified by the model. This technique can be more effective than other techniques because it does not require knowledge of the model's parameters or the gradients of the model. DeepFool attack is an example of such an attack.
\textit{\textbf{Figure 1 represents the three white-box attacks used in this study.}}

\begin{figure}[h!]
\centering
{\includegraphics[scale=1.2]{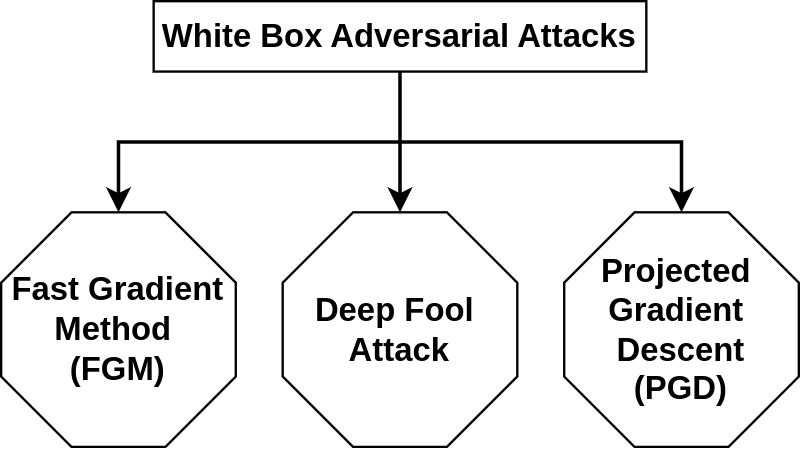}}
\caption{Three different types of white box adversarial attacks analyzed in this study}
\label{fig:sim1}
\end{figure}

\section{Method/ Framework}
\subsection{Original Data Images}
 Various research has been done on the automated classification of histopathological cancer using different datasets. For our experiment, we used the benchmark dataset known as PatchCamelyon(PCam) \cite{veeling2018rotation}. This is a large-scale patch-level data set derived from Camelyon16 \cite{bejnordi2017diagnostic} data.

\begin{figure}[h!]
    \centering
        \includegraphics[scale=1.85]{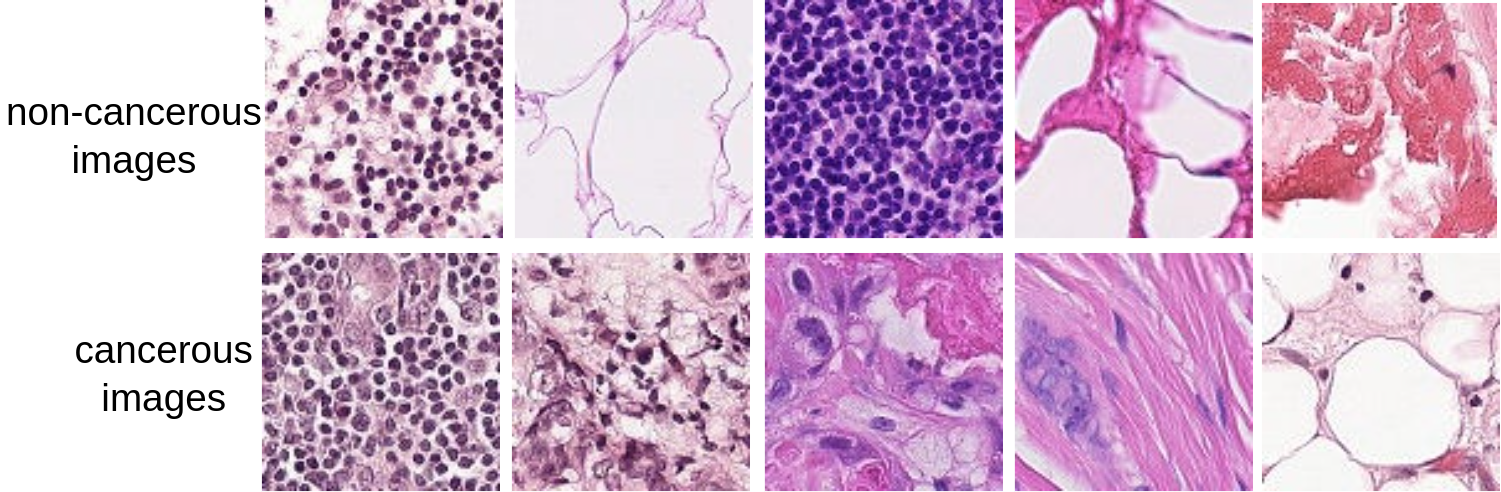}
    \caption{Data samples: non-cancerous and cancerous images}
    \label{fig:sampImages}
\end{figure}

  The aggregate of the patches makes up the slide-level image, which can be used to predict the likelihood of metastases, stage cancer. Example of patch data samples showing likelihood of cancer is shown in Fig.~\ref{fig:sampImages}. The data set contains total 327,680 images. However, we used 10k images in this work model. 
\FloatBarrier


\subsection{Classical and Hybrid Classical-Quantum Binary Image Classification Models}
Convolutional Neural Networks(CNN) are widely used in image-related operations due to their formidable performance. Instead of designing and training neural networks from scratch, different pre-trained transfer learning models \cite{transferLearning1} are used to enhance image classification performance. In our experiment, we  used well-known transfer learning models like VGG16 \cite{vgg16}, InceptionV3 \cite{inception_v3}, Resnet18 \cite{resnet18Org} and Alexnet \cite{alexnetOrg}. These models are trained on an ImageNet dataset with 1000 target categories. However, initial layers of pre-trained models can act as feature extraction layers for customizing image classification tasks for the newer datasets. In our experiment, we have designed the classical and hybrid classical-quantum neural network using the pre-trained transfer learning models mentioned above. These neural networks are fine-tuned by replacing the final fully connected layer with the classical or quantum layer while keeping the weights of initial layers constant for feature extraction. 


For the classical model, `N' features are extracted from the input image using initial layers of a specific pre-trained transfer learning model. These features are inputted into the hidden layer, which consists of a fully connected layer of `N' neurons with activation functions like ReLU, or sigmoid. Finally, an output layer is introduced with neurons equal to the number of target classes in our study, equal to two. This provides us with a comparable architecture with classical-quantum models since we introduce our quantum layers between the fully connected layers of the hidden layer and output layer.
\begin{figure}[htpb]
    \centering
    \includegraphics[scale=1.5]{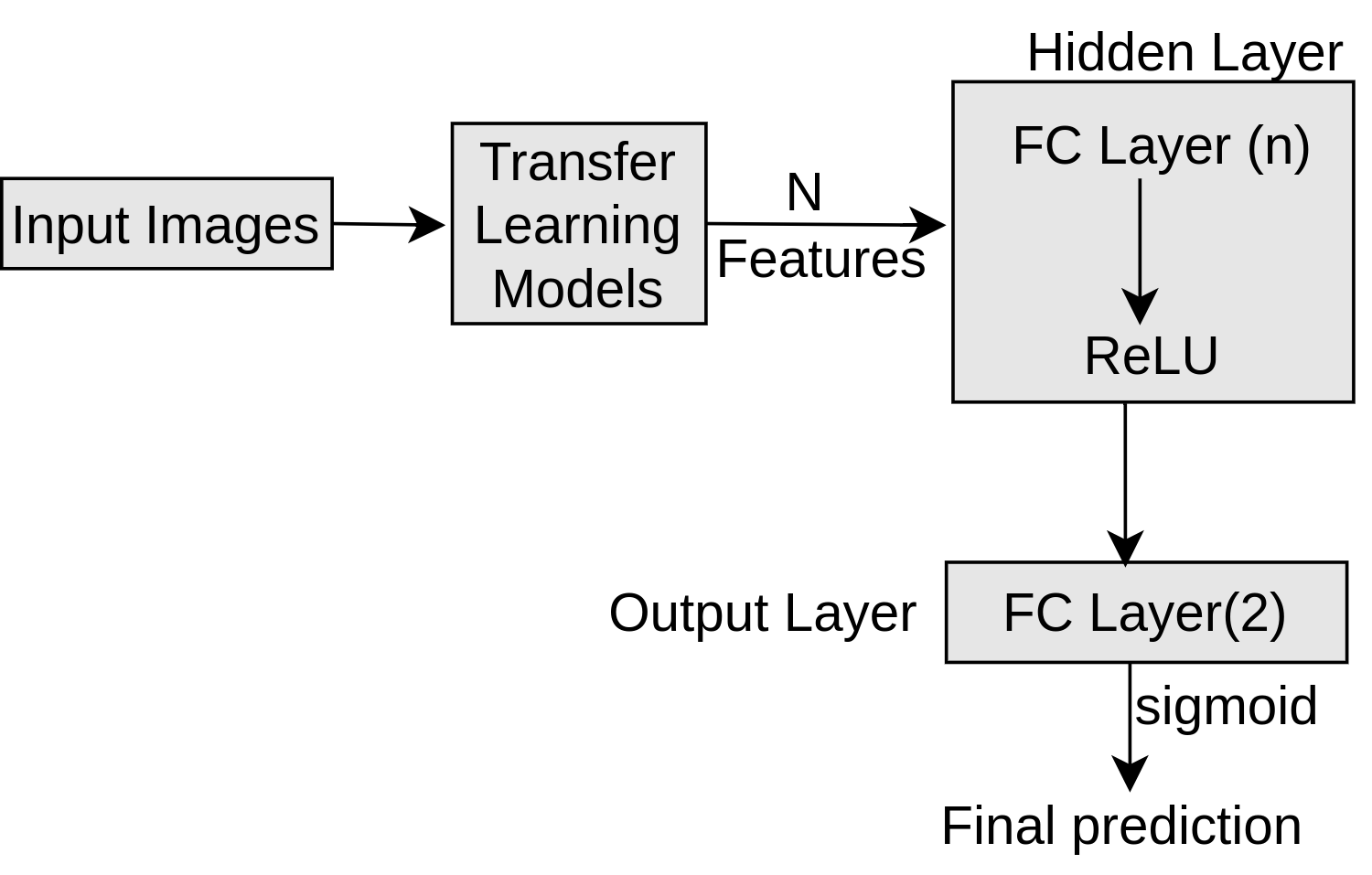}
    \caption{Model Architecture: Classical}
    \label{fig:classical_str}
\end{figure}

For the hybrid classical-quantum model, a Quantum Neural Network (QNN) layer  based on variational quantum circuits (VQC) is sandwiched between two classical neural network layers \cite{xanaduQTL}. Features extracted from the initial layers are thresholded  between 2 to 8 since an equivalent number of n-qubit systems is initialized for the quantum layer, and features are embedded in the quantum systems. This is done keeping in consideration of various quantum hardware constraint.
These features are the same in number as the number of qubits used in VQC and are the inputs to the QNN layer. For the hybrid classical-quantum model, a Quantum Neural Network (QNN) layer  based on variational quantum circuits (VQC) is sandwiched between two classical neural network layers. The outputs from the QNN layer are input to the final output layer.  
\begin{figure}[h!]
    \centering
    \includegraphics[scale=1.2]{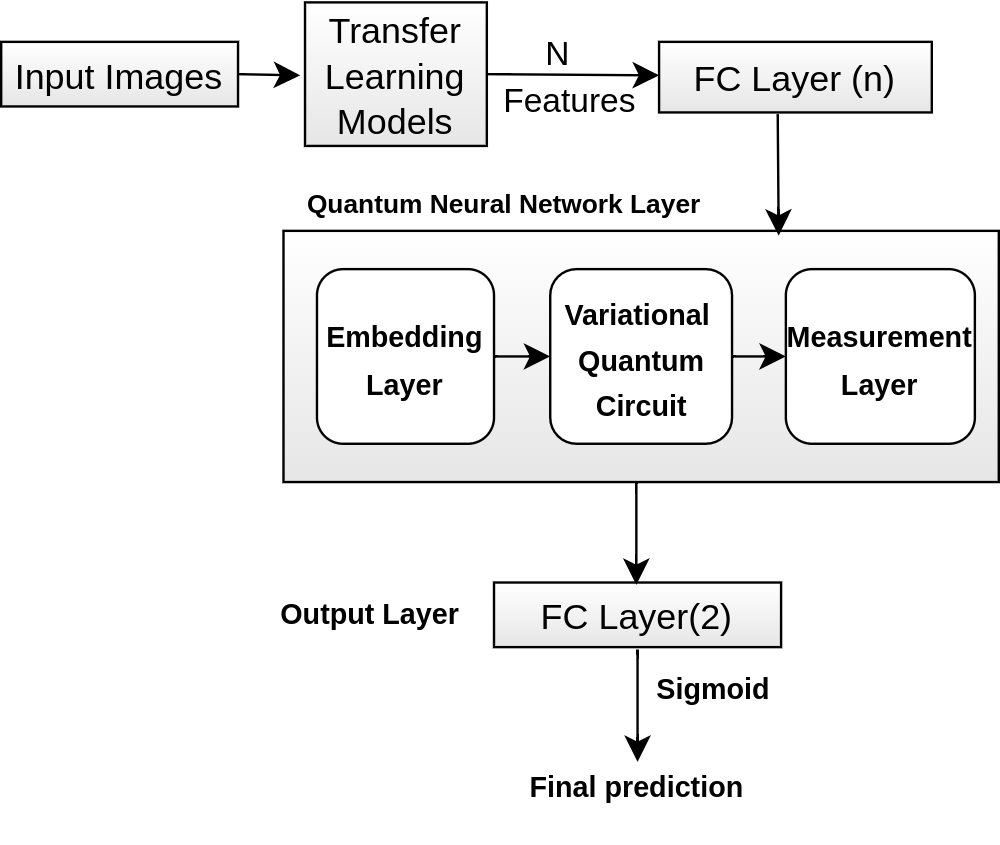}
    \caption{Model Architecture: Hybrid Classical-Quantum}
    \label{fig:hybrid_str}
\end{figure}

From the model architecture shown in Fig.\ref{fig:hybrid_str}, we can observe that the quantum operation is performed on three different layers of QNN layer. The first layer is the Embedding layer which is responsible for mapping the data in the classical vector into a quantum state. In order to map classical data into the quantum state, different single qubit gates like Hadamard gate, U1, U2, U3 gates, Rotational X, Rotational Y, and Rotational Z gates are used. Next layer is the variational quantum circuit layer which is the concatenation of quantum layers of depth `d'. Two qubit gates like controlled Z, controlled NOT, controlled RX are used with parameterized single qubit gates to design parameterized circuits. The next step is to map the obtained outputs from the quantum circuit to the classical domain. For this the expectation values from the quantum circuit is measured on one of X, Y or Z basis. The result from this layer is the input to the next classical layer which is the output layer. In our case, the output layer is the fully connected layer with two neurons for binary classification with sigmoid activation. 
\begin{figure}[h!]
    \centering
    \includegraphics[scale=0.155]{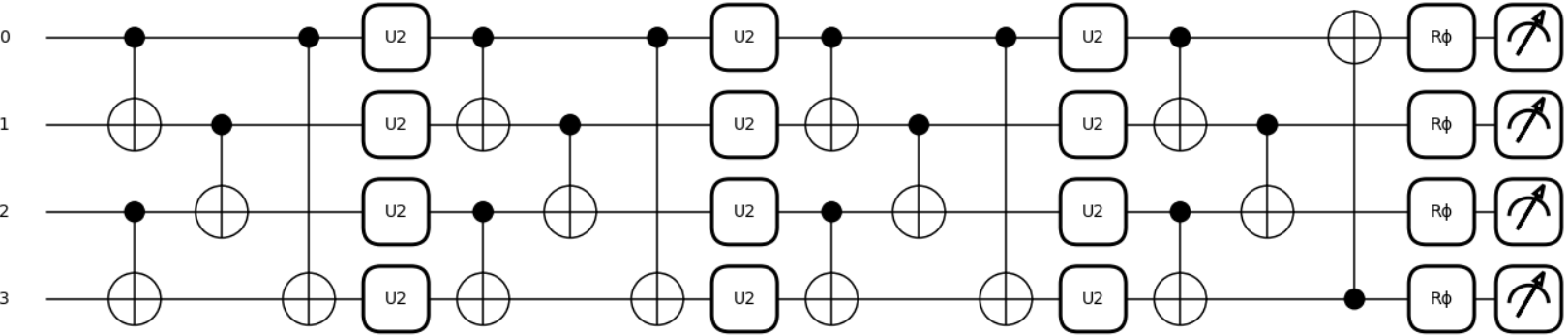}
    \caption{VQC-1 used in our hybrid classical-quantum model which achieved highest classification accuracy on 1000 subsets of data}
    \label{fig:circuit_1}
\end{figure}

\begin{figure}[h!]
    \centering
    \includegraphics[scale=0.154]{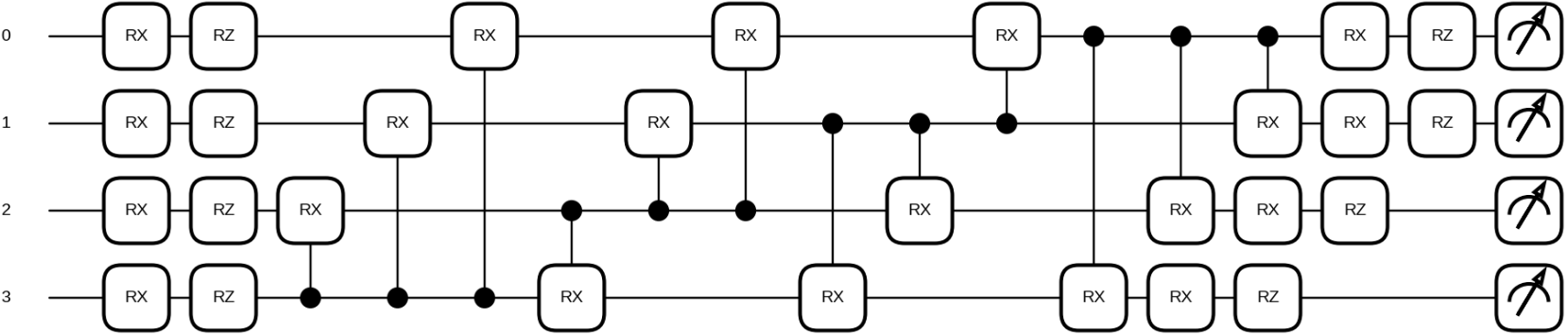}
    \caption{VQC-6 used in our hybrid classical-quantum model which achieved highest classification accuracy on 10000 subsets of data}
    \label{fig:circuit_6}
\end{figure}

In our experiment, we have used pre-trained transfer learning models from torchvision \cite{torchvision} and designed a classification model using Pytorch \cite{pytorch}. For hybrid classical-quantum model the circuit is designed using PennyLane \cite{Pennylane} and the integration of quantum node with classical PyTorch layer is done using TorchLayer class of qnn module from PennyLane. The quantum circuits are executed on the pennylane default simulator.

\subsection{Preparing Adversarial Images}
\begin{figure}[h!]
    \centering
    \includegraphics[scale=0.8]{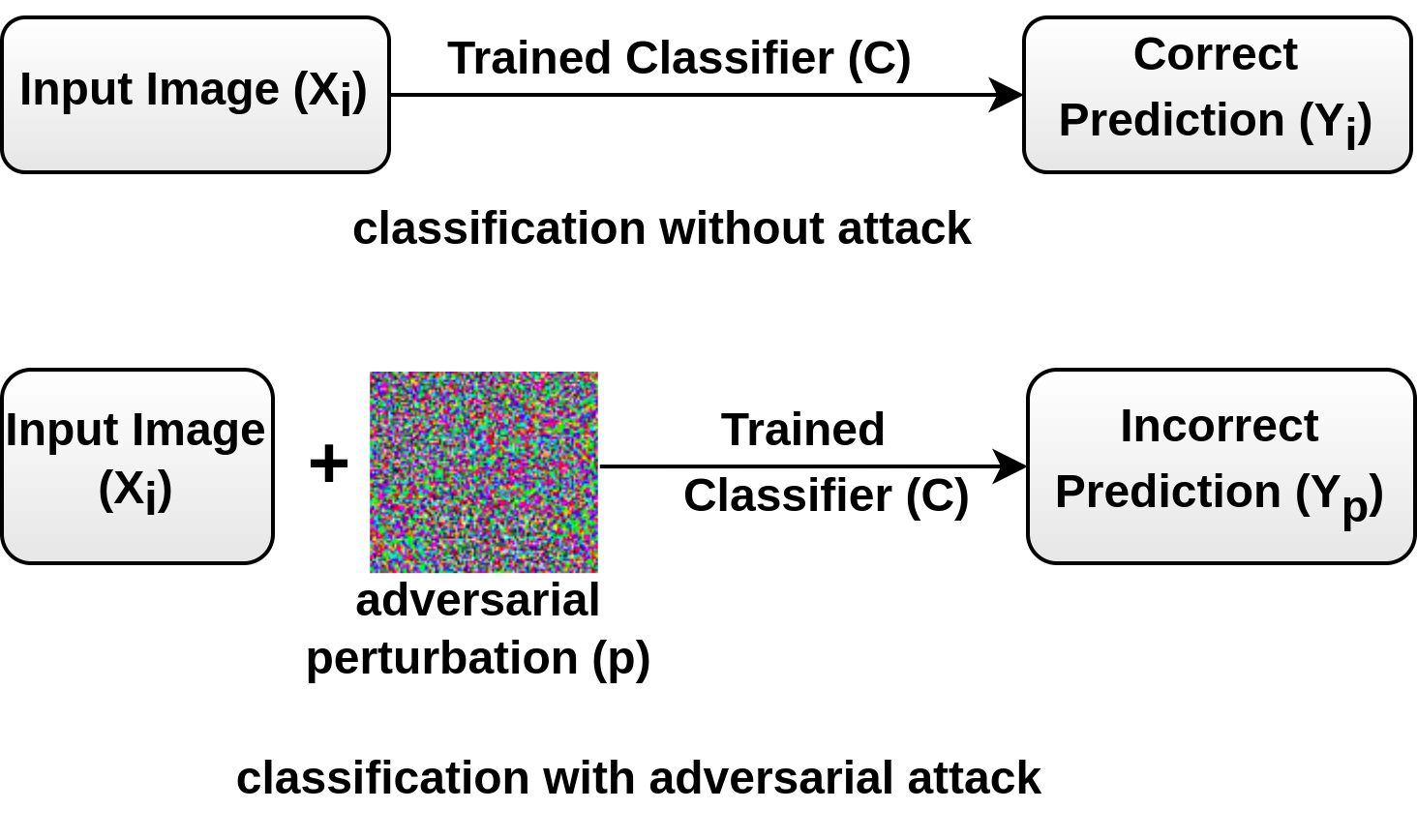}
    \caption{Adversarial Attack Scenario: $X_{i}$ is the input image, $Y_{i}$ is the prediction without attack and ${Y_p}$ is the prediction with adversarial perturbtaion}
    \label{fig:adversarial_attack_scenario}
\end{figure}

Deep learning models based on CNN have achieved higher accuracy in histopathological cancer detection \cite{veeling2018rotation} \cite{bejnordi2017diagnostic}. However, these classification models are highly vulnerable to different kinds of adversarial attacks which leads to unexpected predictions with higher confidence scores. The analysis of adversarial attacks on such models helps to better estimate the reliability of the classifier model and design a method to defend against such attacks. The general scenario of an adversarial attack on image classification model is depicted in figure \ref{fig:adversarial_attack_scenario}.

Under normal conditions, the input image $X_i$ is fed into the classifier $C$, which gives output $Y_i$ which is the predicted target class corresponding to the input sample $X_i$. Under an adversarial attack, the input sample is intentionally adultered with random noise which is commonly known as adversarial perturbations. The integration of noise with the images are human-unobtrusive but they lead the classifier to misclassify the input sample with a high confidence score. 

In this work model, we have evaluated the performance of different classical and hybrid classical-quantum models under three types of white-box adversarial attacks: Fast Gradient Sign Method(FGSM) attack, Deep Fool attack and Projected Gradient Descent (PGD) attack. For the generation of adversarial images, we have used an untargeted attack method. In FGSM attack, adversarial images are generated using the sign of the gradient \cite{goodfellow2014explaining}. Input images are fed into classifier to generate the prediction and loss. Then the gradient of loss is calculated with respect to input. The gradient undergoes sign function to calculate its sign value. The process of generating adversarial perturbation using FGSM is expressed in equation \ref{eqn:fgsm_eqn}. 
\begin{equation}
    \centering
    X_{adv}= X + \epsilon * sign(\nabla_x\mathcal{L}(C,X,Y))
    \label{eqn:fgsm_eqn}
\end{equation}
where,
\begin{itemize}
    \item [] $X_{adv}$ = the generated adversarial image
    \item [] $\epsilon$ = perturbation coefficient which is lower enough to detect from human eye, higher enough to fool the classifier
    \item [] $\mathcal{L}$= loss function for classifier $C$ with input $X$ and target $Y$
\end{itemize}
PGD attack is another variant of gradient-based attack. Perturbations are generated by running FGSM multiple time with small step size and the adversarial values are clipped after each step to the perturbation constraint which are already defined \cite{PGD1} \cite{PGD2}. The DeepFool attack algorithm works on the basis of decision boundaries to generate perturbations \cite{DeepFoolOriginal}. The algorithm iteratively computes the gradient of the classification model's output with respect to the input image and then determines the direction of the gradient that leads to the smallest change in the image classification. This process is repeated until there is a change in the prediction label of the input or the current iteration is the maximum.

\begin{figure}[h!]
    \centering
    \includegraphics[scale=7]{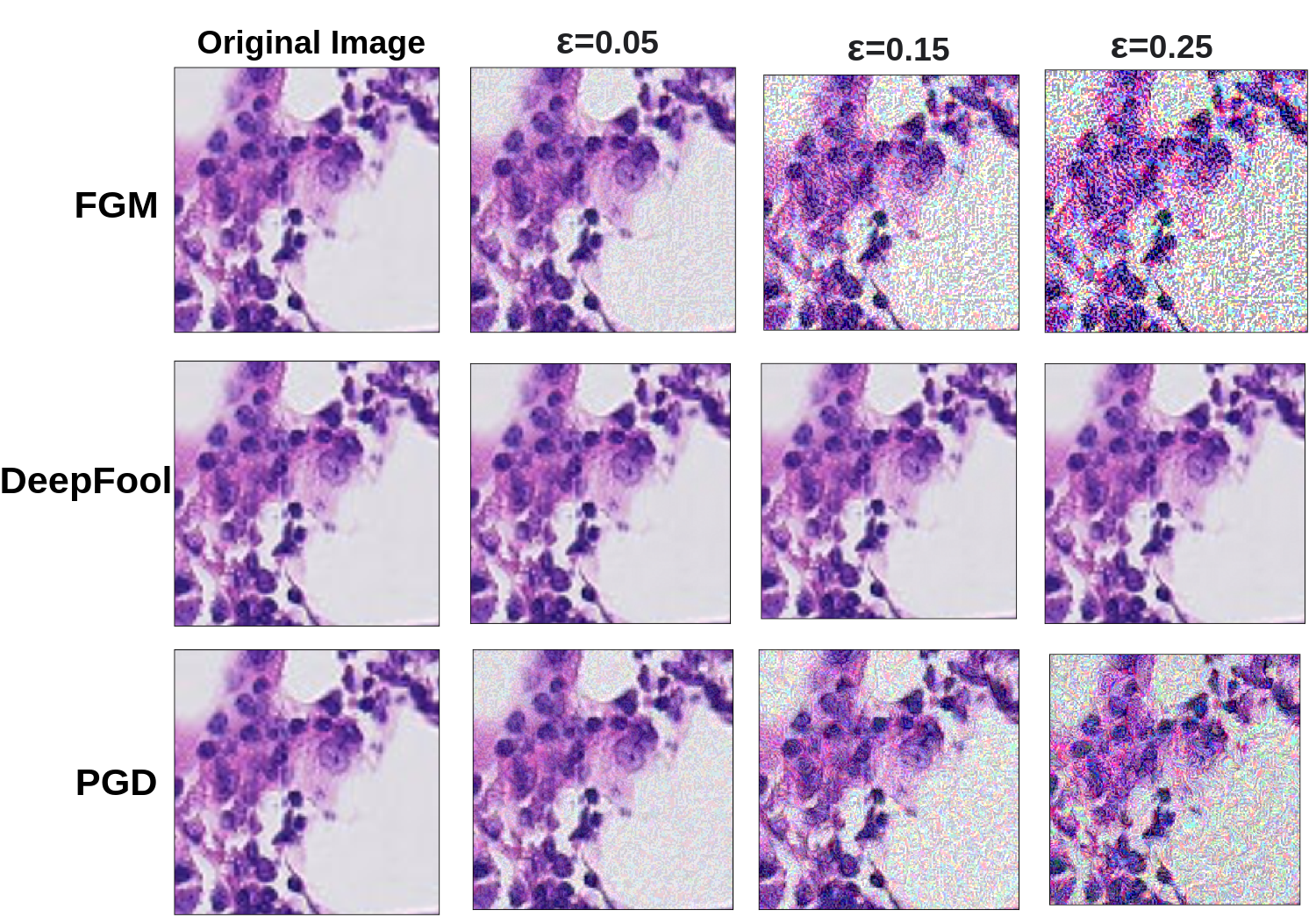}
    \caption{Adversarial Images: each row contains different types of attacks and each column contains perturbed images under different values of epsilon($\epsilon$)}
    \label{fig:adversarial_all_figs}
\end{figure}

Figure \ref{fig:adversarial_all_figs} shows a sample of adversarial images under different values of epsilon($\epsilon$).`$\epsilon$' is the perturbation coefficient used in each of the adversarial attack algorithms.

\section{Results}
\begin{table*}[htpb]
\centering
\caption{Model Performance with each of the Adversarial Attacks used in the Study}
\begin{tabular}{|p{1.21 cm}|p{1.1 cm}|p{0.86 cm}|p{1 cm}|p{0.6 cm}|p{0.8 cm}|p{0.8 cm}|p{0.7 cm}|p{1.3 cm}|p{1.3 cm}|p{1.3 cm}|}
\hline
\textbf{Model} & \textbf{Comp. type}&\textbf{no. of Images} &  \textbf{Accuracy on Simulator}  & \textbf{VQC} & \textbf{Express- ibility}  & \textbf{no. of Qubits} & \textbf{epsilon ($\epsilon$)} & \textbf{Accuracy under FGM Attack} & \textbf{Accuracy under Deep Fool Attack} & \textbf{Accuracy under PGD Attack} \\ \hline

Classical CNN & Classical  & 5000 &  76.40  & N/A & N/A & N/A  & 0.05 \newline 0.15 \newline 0.25 & 52.90\newline53.80\newline50.90 & 28.49\newline28.90\newline30.00 & 61.30\newline28.59\newline23.70 \\ \hline  

Classical CNN2 & Classical & 10000 & 80.95  & N/A & N/A & N/A & 0.05\newline0.15\newline0.25 & 30.40\newline34.20\newline50.20 & 42.75\newline43.40\newline44.00 & 39.35\newline22.00\newline22.55 \\ \hline 

Classical ResNet18 & Classical & 1000 & 86.00  & N/A & N/A & N/A  & 0.05 \newline 0.15 \newline 0.25 & 31.00\newline27.00\newline34.50 & 17.00\newline16.50\newline16.50 & 30.00\newline14.00\newline14.00 \\ \hline

Classical ResNet18 & Classical & 10000 & 89.50  & N/A & N/A & N/A  & 0.05 \newline 0.15 \newline 0.25 & 19.70\newline16.55\newline17.90 & 16.25\newline17.29\newline18.15 & 38.35\newline10.54\newline10.50 \\ \hline


Hybrid ResNet18 & Hybrid Classical Quantum & 1000 & 88.50  & 1 & 1.431& 4 & 0.05 \newline 0.15 \newline 0.25 & 46.50\newline60.50\newline65.50 & 38.00\newline39.00\newline39.50 & 50.50\newline50.00\newline50.00\\ \hline

Hybrid VGG 16 & Hybrid Classical Quantum & 1000 & 79.50 & 2 & 1.078 & 5  & 0.05 \newline 0.15 \newline 0.25 & 44.50\newline41.00\newline42.00 & 21.00\newline22.50\newline23.00 & 43.00\newline48.00\newline53.00 \\ \hline 

Hybrid \newline Inceptionv3 & Hybrid Classical Quantum & 1000 & 77.00  & 3 & 1.007& 5 & 0.05 \newline 0.15 \newline 0.25 & 59.00\newline50.00\newline50.00 & 40.00\newline42.00\newline43.50 & 53.00\newline48.00\newline49.50\\ \hline

Hybrid AlexNet & Hybrid Classical Quantum & 1000 & 80.00  & 4 & 0.201& 5 & 0.05 \newline 0.15 \newline 0.25 & 45.00\newline56.99\newline47.00 & 28.00\newline28.99\newline33.50 & 51.50\newline50.00\newline50.00\\ \hline

Hybrid ResNet18 & Hybrid Classical Quantum & 5000 & 83.80  & 5 & 1.201 & 7 & 0.05 \newline 0.15 \newline 0.25 & 46.40\newline58.50\newline68.60 & 17.20\newline20.00\newline22.10 & 44.10\newline41.30\newline34.00 \\ \hline

Hybrid ResNet18 & Hybrid Classical Quantum & 10000 & 82.35  & 1 & 1.431& 4 & 0.05 \newline 0.15 \newline 0.25 & 50.94\newline69.69\newline77.75 & 46.00\newline47.80\newline48.80 & 52.70\newline49.95\newline50.05 \\ \hline

HQC ResNet18  & Hybrid Classical Quantum & 10000 & 84.30  & 6 & 0.011& 4 & 0.05\newline0.15\newline0.25 & 54.05\newline63.55\newline62.35 & 16.50\newline20.34\newline24.45 & 55.65\newline45.55\newline45.35\\ \hline

\end{tabular}
\label{tab:results}
\end{table*}

We evaluated the performance of different classical and hybrid classical-quantum binary image classification models with and without adversarial perturbations. For classical computation, we created 4 models out of which 2 are custom-defined convolutional neural networks and the other two are resnet18 transfer learning based models. To generate adversarial perturbation using hybrid classical-quantum transfer learning models, we used 4 widely used pre-trained transfer learning models: VGG16, Resnet18, Alexnet and Inceptionv3. We created and trained 7 hybrid classical-quantum models with 6 different VQCs. These classical and hybrid models are trained and tested on different subsets of data.

Table \ref{tab:results} outlines the performance of different classical and hybrid classical-quantum models from our experiment. Column I represents the name of models from our experiment. The computation type of different model architectures are mentioned in column II. The computations are either classical or hybrid classical-quantum. Column III represents the number of images included in the subset of data. We used 3 subsets of data with 1000, 5000 and 10000 images. For the subset with 1000 images, we splitted it into 80:20 ratio of train and test sets. For datasets with higher number of images we splitted them into 60:20:20  ratio of train, validation and test sets. The test accuracy achieved from each of these models without adversarial perturbations is included in column IV. Column V includes the VQCs used in our hybrid classical-quantum transfer learning models. Expressibility \cite{expressibilityPaper} and the number of qubits of VQC in column V are depicted in columns VI and VII respectively. Different values of perturbation coefficients($\epsilon$) used in each of the attack models are included in column VIII. For our experiment, we evaluated the performance for each of the classical and hybrid classical-quantum models under 3 perturbation coefficients: 0.05, 0.15 and 0.25. The accuracy of each model tested on adversarial samples generated using FGM, DeepFool and PGD with perturbation coefficient ($\epsilon$) are listed in columns IX, X and XI respectively.

\subsection{Experimental results of classical models}
From the table \ref{tab:results} we can see that in classical models without adversarial perturbation, the classification accuracies of CNN models with transfer learning are higher than the classical models without transfer learning. The highest classification accuracy achieved from the classical models is 89.5 percent which is resnet18 transfer learning-based model trained on a subset of data with 10000 images. Whereas, in the case of evaluating classical models with adversarial perturbations the models trained without using pre-trained transfer learning models have achieved higher classification accuracies. Under the adversarial perturbation generated using FGM attack the highest classification accuracy is 52.9 percent which is obtained from the classical CNN model trained on subset of 5000 datasets. Using deefool model to generate perturbations the highest classification accuracy of 44.0 percent is achieved from the classical CNN model trained on a subset of 10000 images. Under PGD attack, the classical model trained on subset of 5000 images achieved the highest classification accuracy which is 61.3 percent.

\subsection{Experimental results of Hybrid classical-quantum models}
For hybrid classical-quantum models we evaluated the performance of each transfer learning models with small subset of images (1000 images). The model with highest classification accuracy is selected to train with higher number of images. In our experiment, Resnet18 transfer learning model outperformed other 3 transfer learning models. The hybrid model trained with 1000 subset of images achieved highest classification accuracy which is found to be 88.50 percent without adversarial perturbations. For the subset of images with 10000 images without adversarial perturbations the highest classification accuracy is  84.30 percent. 77.75 percent of classification accuracy under FGM attack is achieved from hybrid classical-quantum model with VQC-1. The hybrid classical-quantum model based of resnet18 transfer learning model with VQC-1 outperformed other hybrid models under FGM and DeepFool adversarial attacks. Under FGM and DeepFool attacks classification accuracies achieved are 77.75($\epsilon$=0.25) and 48.80($\epsilon$=0.25) respectively. Under PGD attack hybrid classical-quantum model with VQC-6 achieved highest classification accuracy which is found to be 55.65($\epsilon$=0.05).

\section{Conclusion and Future Work}
Machine learning models have achieved state-of-the-art performance on different medical image-related operations like image classification, image segmentation. However, the use of these models in medical sectors are extremely vulnerable to different kinds of malicious attacks commonly known as adversarial attacks.
In this experiment, we explored the impact of adversarial attacks on different classical and hybrid classical-quantum image classification models for histopathological cancer detection. For the evaluation of classical models, we chose 4 classical models with varying subsets of images. Similarly, we chose 7 hybrid classical-quantum models to evaluate their performance under different adversarial attacks. The experiment we performed shows that both the classical and hybrid classical-quantum models deployed are highly vulnerable to adversarial attacks. However, the success rate of defence against such adversarial perturbations of hybrid classical-quantum models are higher than that of classical classification models. 

Currently, we have performed our experiment on a quantum default simulator from PennyLane. As a consequent work, we plan to test the impact of different adversarial attacks on real quantum hardware. The current experiment shows that there is a potential to develop resilient models towards different kinds of malicious attacks using the architecture of hybrid classical-quantum models.


\bibliographystyle{ieeetr}
\bibliography{main}
\end{document}